\documentclass[reprint,twocolumn,superscriptaddress,pre,showpacs,amsmath,amssymb,aps]{revtex4-1}

\usepackage{natbib}
\usepackage{graphicx,color,rotating}
\usepackage[latin1]{inputenc}
\usepackage{textcomp}
\usepackage{dcolumn}
\usepackage{bm}
\usepackage{units}    
\usepackage{subfigure}

\newcommand{\notop}{{{}_{}}}

\renewcommand{\vec}[1]{\bm{#1}}
\newcommand{\ee}{\mathrm{e}}
\newcommand{\ii}{\mathrm{i}}
\newcommand{\dm}{\mathrm{d}}

\DeclareMathOperator{\re}{Re}

\newcommand{\iot}{{\ii\omega t}}

\newcommand{\nablabf}{\boldsymbol{\nabla}}
\newcommand{\nablabfTi}{\tilde{\nablabf}}

\newcommand{\Lapl}{\nabla^2}

\newcommand{\eee}{\vec{e}}
\newcommand{\een}{\vec{e}^\notop}

\newcommand{\FFF}{\vec{F}}

\newcommand{\FFFrad}{\vec{F}^{\mathrm{rad}}}

\newcommand{\FFFradExt}{\vec{F}^{\mathstrut\mathrm{rad}}_{\mathstrut\mathrm{ext}}}
\newcommand{\FFFradInt}{\vec{F}^{\mathrm{rad}}_{\mathrm{int}}}

\newcommand{\rrr}{\vec{r}}
\newcommand{\rrn}{\rrr^\notop}

\newcommand{\vvv}{\vec{v}}

\newcommand{\vvvIn}{\vec{v}^\notop_\mathrm{in}}

\newcommand{\zerovec}{\boldsymbol{0}}

\newcommand{\calO}{\mathcal{O}}

\newcommand{\calS}{\mathcal{S}^\notop}

\newcommand{\calU}{\mathcal{U}}

\newcommand{\kapTi}{\tilde{\kappa}}

\newcommand{\UpsO}{U^{{}}_0}
\newcommand{\UpsI}{U^{{}}_1}

\newcommand{\epss}{\epsilon_s}
\newcommand{\epsp}{\epsilon_p}

\newcommand{\cO}{c^{{}}_0}
\newcommand{\cOsqr}{c^{\,2_{}}_0}

\newcommand{\kapO}{\kappa^{{}}_0}

\newcommand{\pInUconj}[1]{\big[p^{(0)_{}}_\mathrm{in}\big]^*}

\newcommand{\phiIn}{\phi^{{}}_{\mathstrut\mathrm{in}}}

\newcommand{\phiExt}{\phi^{{}}_{\mathstrut\mathrm{ext}}}
\newcommand{\phiSExt}{\phi^{*_{}}_{\mathstrut\mathrm{ext}}}

\newcommand{\phiMsc}{\phi^{{}}_{\mathstrut\mathrm{msc}}}

\newcommand{\phiSc}{\phi^{{}}_\mathrm{sc}}

\newcommand{\rhoO}{\rho^\notop_0}

\newcommand{\rhoIn}{\rho^\notop_\mathrm{in}}

\newcommand{\rhoTi}{\tilde{\rho}}

\newcommand{\SIm}{\textrm{m}}

\newcommand{\SImum}{\textrm{\textmu{}m}}

\newcommand{\beq}[1]{\begin{equation} \eqlab{#1}}
\newcommand{\eeq}{\end{equation}}
\newcommand{\bsub}{\begin{subequations}}
\newcommand{\esub}{\end{subequations}}
\def\bal#1\eal{\begin{align}#1\end{align}}
\def\bsubal#1\esubal{\bsub \begin{align}#1\end{align} \esub}

\newcommand{\eqlab}[1]{\label{eq:#1}}
\renewcommand{\eqref}[1]{Eq.~(\ref{eq:#1})}
\newcommand{\eqsref}[2]{Eqs.~(\ref{eq:#1}) and~(\ref{eq:#2})}

\newcommand{\figref}[1]{Fig.~\ref{fig:#1}}

\newcommand{\figlab}[1]{\label{fig:#1}}

\newcommand{\seclab}[1]{\label{sec:#1}}

\begin{document}

\title{Acoustic interaction forces between small particles in an ideal fluid}

\author{Glauber T. Silva}
\affiliation{Physical Acoustics Group, Instituto de F\'isica, Universidade Federal de Alagoas,\\
Macei\'o, AL 57072-970, Brazil}
\email{glauber@pq.cnpq.br}

\author{Henrik Bruus}
\affiliation{Department of Physics, Technical University of Denmark,
DTU Physics Building 309, DK-2800 Kongens Lyngby, Denmark}
\email{bruus@fysik.dtu.dk}


\begin{abstract}
\vspace*{-5mm}
\centerline{\small (Submitted to Phys.\ Rev.\ E, 24 August 2014)}

\vspace*{5mm}
We present a theoretical expression for the acoustic interaction force between small spherical particles suspended in an ideal fluid exposed to an external acoustic wave. The acoustic interaction force is the part of the acoustic radiation force on one given particle involving the scattered waves from the other particles. The particles, either compressible liquid droplets or elastic microspheres, are considered to be much smaller than the acoustic wavelength. In this so-called Rayleigh limit, the acoustic interaction forces between the particles are well approximated by gradients of pair-interaction potentials with no restriction on the inter-particle distance. The theory is applied to studies of the acoustic interaction force on a particle suspension in either standing or traveling plane waves. The results show aggregation regions along the wave propagation direction, while particles may attract or repel each other in the transverse direction. In addition, a mean-field approximation is developed to
describe the acoustic interaction force in an emulsion of oil droplets in water.
\end{abstract}



\maketitle


\section{Introduction}

Techniques relying on acoustofluidic forces, such as acoustic radiation force and streaming, are currently used in many different ways to handle suspended cells, microparticles and fluids non-intrusively and label-free in microfluidic setups such as separation, trapping, and sorting of cells, particle manipulation, as well as generation and control of fluid motion~\cite{Laurell2007, Bruus2011c, Ding2012}. Experimentally, ultrasound waves emitted into a particle suspension give rise to acoustic streaming of the carrier fluid, and are responsible for the two acoustofluidic forces driving the acoustophoretic motion of the suspended particles: the acoustic radiation force and the Stokes drag force from acoustic streaming. The theoretical description of these complex, non-linear acoustic effects is not yet complete, and in this paper we develop the theory of the acoustic radiation force, which dominates the motion of the larger microparticles~\cite{Barnkob2012a}.

Concerning the radiation force exerted on a single particle, the so-called primary radiation force $\FFFrad$, recent studies by Doinikov~\cite{Doinikov1997}, Danilov and Mironov~\cite{Danilov2000}, as well as Settnes and Bruus~\cite{Settnes2012} have advanced the theoretical treatment beyond the seminal contributions by King~\cite{King1934}, Yosioka and Kawasima~\cite{Yosioka1955}, and Gorkov~\cite{Gorkov1962}. The main improvement found in these recent studies is the introduction of thermoviscous effects in both the incident ultrasound waves and in the scattered wave from the particle. However, in a particle suspension exposed to an external acoustic wave, a secondary radiation force appears, the so-called acoustic interaction force $\FFFradInt$. For a given particle, the acoustic interaction force is caused by the scattered waves from the other particles. Investigations on this force dates back to the nineteenth century, when Bjerknes studied the mutual force between a pair of bubbles~\cite{Bjerknes1906},
and the analysis performed by K\"onig on the acoustic interaction force between two rigid spheres~\cite{Konig1891}. Subsequently, this force was investigated considering short-range interaction between particles of the types rigid-rigid~\cite{Emblenton1962, Nyborg1989}, bubble-bubble~\cite{Doinikov1995, Crum1975}, bubble-rigid~\cite{Doinikov1996}, and bubble-droplet~\cite{Doinikov1996a};
whereas long-range rigid-rigid~\cite{Zhuk1985} and
bubble-bubble~\cite{Doinikov1999, Doinikov2002} interactions have also been studied. The acoustic interaction force between two droplets aligned relative to an incident plane wave with arbitrary inter-particle distance
was also analyzed~\cite{Zheng1995}. Moreover, bubble-bubble interaction at any separation distance has also been analyzed through a semi-numerical scheme based on the partial-wave expansion method and the translational addition theorem of spherical functions~\cite{Doinikov2001}.

The current literature on the acoustic interaction force lacks an investigation on a suspension composed of compressional fluid droplets or solid elastic particles without any restriction on the inter-particle distances. These kind of particles are often used in experiments on acoustofluidics, acoustical tweezers, and demulsification of particle-water mixtures by ultrasound. It is our goal here to provide an analytical expression for the acoustic interaction force between suspended droplets or solid elastic microparticles in an inviscid fluid. The proposed method, which takes the form of a scalar potential theory for the acoustic interaction force, extends the single-particle radiation force theory developed by Gorkov~\cite{Gorkov1962} to include re-scattering events between particles in the suspension. The method is applied to various examples of the acoustic interaction force in the case of either a standing or a traveling external plane wave, and a mean-field theory is proposed and applied to compute the
acoustic interaction force between the drops in an emulsion of oil drops in water.

\section{Theory}
\seclab{theory}
The linear wave theory for the acoustic fields in an unbounded, isotropic fluid of density $\rhoO$ and isentropic compressibility $\kappa_{0}= 1/(\rhoO\cOsqr)$, where $\cO$ is the adiabatic sound velocity in the fluid, is standard textbook material~\cite{Morse1986, Pierce1989, Blackstock2000}.
We neglect the viscous dissipation of the acoustic field in the particle suspension, which is a good approximation for particle radii much larger than the width of the viscous boundary layer~\cite{Settnes2012} and for frequencies much lower than hypersound frequencies (below GHz for water).
Consequently,  a time-harmonic acoustic wave can be described by the velocity potential $\Phi(\rrr,t)$, where $\rrr$ is postion and $t$ is time, in terms of a complex-valued phase factor $\ee^{-\iot}$, where $\omega$ is the angular wave frequency, and an amplitude function $\phi(\rrr)$, which satisfies the Helmholtz wave equation,
%
\bsubal
 \eqlab{HarmonicPhase}
 \Phi(\rrr,t) &= \phi(\rrr)\:\ee^{-\iot},\\
 \eqlab{Helmholtz}
 \Lapl \phi(\rrr) &= -k^2\:\phi(\rrr), \text{ with } k = \frac{\omega}{\cO}.
 \esubal
In terms of the potential $\phi(\rrr)$, the amplitude function of the pressure $p(\rrr)$, the density $\rho(\rrr)$, and the velocity $\vvv(\rrr)$ are given by
 \bsubal
 \eqlab{pphi}
 p(\rrr)    &= \ii\omega\rhoO\:\phi(\rrr),\\
 \eqlab{rhophi}
 \rho(\rrr) &= \ii\frac{\omega\rhoO}{\cOsqr}\:\phi(\rrr),\\
 \eqlab{vphi}
 \vvv(\rrr) &= \nablabf\phi(\rrr).
 \esubal

In the following we outline some fundamental concepts of acoustic scattering and radiation forces on small particles suspended in the fluid.

\subsection{Single-particle scattering in the Rayleigh limit}
\label{rayleigh_sca}
Consider a monochromatic acoustic wave represented by the velocity potential amplitude $\phiIn(\rrr)$ incident on and scattering off a small spherical particle suspended in the medium. The scattered wave adds to acoustic wave incident on any other particle in the suspension, so the first particle acts as a source of additional acoustic radiation forces felt by the other particles in the suspension. All physical quantities related to this source particle are marked by the subscript "s" such as particle radius $a_s$, density $\rho_s$, isentropic compressibility $\kappa_s$, and center position $\rrn_s$, as sketched in \figref{scat_geom}. At any given probe position $\rrn_p$, the outgoing scattered wave from the source particle is represented by the velocity potential amplitude function $\phiSc(\rrn_p|\rrn_s)$, where subscript "p" here and in the following relates to the probe. Throughout this work, we only consider the so-called Rayleigh scattering limit $ka_s \ll 1$. We also assume ideal scattering boundary
conditions, i.e.\ total absorption without reflection of any scattered waves at infinity. In this limit, the acoustic scattering is dominated by the monopole and dipole scattering, and the scattered wave is given by~\cite{Landau1993},
 \begin{align}
 \nonumber
 \phiSc(\rrn_p|\rrn_s)  &=
 \ii f_{0,s} \frac{a_s^3\omega}{3\rho_0}
 \frac{\rhoIn(\rrr_s)\:\ee^{\ii k R_{ps}}}{R_{ps}}\\
 &- f_{1,s} \frac{a_s^3}{2} \nablabf_p \!\cdot\!
 \left[\!\frac{\vvvIn(\rrr_s)\:\ee^{\ii kR_{ps}}}{R_{ps}}\right]
  + \calO\left[\frac{(ka_s)^5}{(kR_{ps})^3}\right],
  \label{phi_sc1}
\end{align}
where $R_{ps}=| \rrn_p - \rrn_s|$, $\nablabf_p$ is nabla acting on $\vec{r}_p$,
and terms of the order $(ka_s)^5/(kR_{ps})^3$ arise from the quadrupolar scattering \cite{Crocker1998}.
The monopole and dipole scattering factors $f_{0,s}$ and $f_{1,s}$
of the source particle are given in terms of the density ratio $\rhoTi_s = \rho_s/\rhoO$ and the
compressibility ratio $\kapTi_s = \kappa_s/\kapO$ as follows \cite{Yosioka1955, Gorkov1962},
 \bsub
 \label{factors}
 \begin{alignat}{2}
 f_{0,s} &= 1 - \kapTi,\\
 f_{1,s} &= \frac{2(\rhoTi_s - 1)}{2\rhoTi_s + 1}.
 \end{alignat}
 \esub

\begin{figure}[t]
\centering
\includegraphics[]{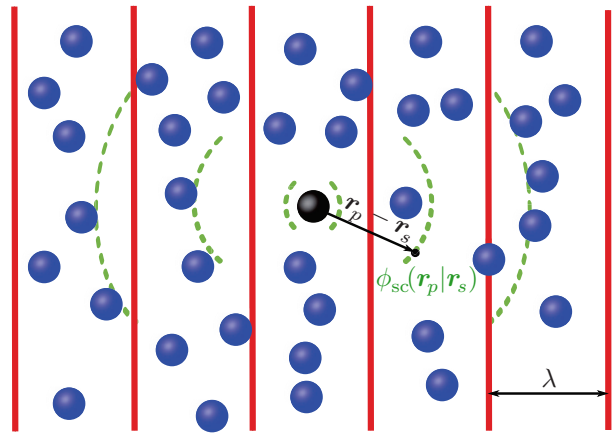}
\caption{\figlab{scat_geom} (Color online)
Sketch of the external incident wave (straight lines) scattering by suspended small spherical particles with radii $a_s\ll \lambda$.
The scattered wave $\phiSc(\rrn_p|\rrn_s)$ from a source particle located at $\rrr_s$
(black sphere),
which is probed at the position $\rrr_p$, is illustrated by dashed arches.}
\end{figure}

For the analysis of the higher-order scattering, it is useful to introduce the scattering parameters $\epss$ and $\epsp$ as well as the dimensionless probe-source distance $x_{ps}$,
 \beq{epsDef}
 \epss = k a_s, \qquad \epsp = k a_p, \qquad x_{ps} = kR_{ps}.
 \eeq
This together with \eqsref{rhophi}{vphi}, can be used to rewrite Eq.~(\ref{phi_sc1}) in terms of a scattering
operator acting on the  incident wave $\phiIn(\rrn_s)$ as
 \begin{align}
 \nonumber
 \phiSc(\rrn_p|\rrn_s) &=   - \epss^3  \frac {\ee^{\ii x_{ps}}}{x_{ps}}
 \Bigg[\frac{f_{0,s}}{3}+ \frac{\ii f_{1,s}}{2}
 \Big(1\!+\!\frac{\ii}{x_{ps}} \Big) \partial_{x_{ps}}  \Bigg]\! \phiIn\!(\rrr_s)\\
 & \qquad +\calO\left(\tilde{\epss}^5\right),
 \label{phi_scatt}
 \end{align}
where $\tilde{\epss} = \epss/x_{ps}^{3/5}$.
Note that $R_{ps} \sim a_s$ implies $x_{ps} \sim \epss$, and thus
$\phiSc(\rrn_p|\rrn_s) =\calO(\epss)$ for probes near the source.

\subsection{Single-particle radiation force}

Once the scattering velocity potential in Eq.~(\ref{phi_scatt}) is known, the resulting acoustic radiation
force acting on a suspended probe particle of radius $a_p$ and scattering coefficients $f_{0,p}$
and $f_{1,p}$ placed  at $\rrr_p$ can be calculated in standard
manners using second-order time-averaged perturbation theory
in the pressure or the particle velocity amplitude~\cite{Gorkov1962, Yosioka1955, Settnes2012}.
In the Rayleigh scattering limit for any incident acoustic wave $\phiIn(\rrr)$,
except plane traveling waves, the radiation force $\FFFrad(\rrn_p)$ is a gradient of a potential $U$ given by
 \bsubal
 \label{Frad}
 \FFFrad(\rrn_p) &= - \nablabf_{\!p} U(\rrr_p),\\
 U(\rrn_p) &=
 -\frac{\epsp^3 \pi \rhoO}{k}
  \bigg[\!\frac{f_{0,p}}{3} \left|\phiIn(\rrn_p)\right|^2 \!-\!
  \frac{f_{1,p}}{2} \left|\nablabfTi_{\!p} \phiIn(\rrn_p) \right|^2 \! \bigg]
  \nonumber
  \\  \eqlab{FradOrder1}
  & \quad + \calO\big(\tilde{\epsp}^5\big),
 \esubal
where $\nablabfTi_{\!p} = (1/k)\nablabf_p$ is the dimensionless nabla operator convenient to use when calculating derivatives of the velocity potential.

\subsection{Scattering in a suspension of particles}
\label{sec:ScatteringSuspension}

We now consider a specific configuration $\calS$ of  $N$ spherical particles arbitrarily placed at the positions
$\rrr_s$ for $s=1, 2, 3, \ldots, N$.
The particle at position $\rrn_s$ has the monopole and dipole
scattering coefficients $f_{0,s}$ and $f_{1,s}$, respectively, as well as radius $a_s$. All particles are
assumed to have  expansion parameters $\epss = ka_s \ll 1$.

An external incident wave with velocity potential
$\phiExt$ hits the $N$-particle suspension and multiple-scattering
processes occurs. The resulting acoustic field $\phiIn(\rrn_p)$ incident at the probe
position $\rrn_p$ can thus be written as
 \beq{phiMscS}
 \phiIn(\rrn_p) = \phiExt(\rrn_p) + \phiMsc(\rrn_p|\calS),
 \eeq
where $\phiMsc(\rrn_p|\calS)$ is that part of the acoustic field at
position $\rrn_p$ that is caused by prior multi-scattering events at one
or more particles in the configuration $\calS$.

In the Rayleigh limit, the multi-scattering contribution to the acoustic wave
$\phiIn(\rrn_p)$ incident at the probe point $\rrn_p$ is dominated by
scattering waves having undergone only a single prior scattering event at some
source point $\rrn_s$ different from $\rrn_p$, that is $s \neq p$. To lowest
scattering order, the multiple-scattering part $\phiMsc(\rrn_p|\calS)$ of the
incident wave at $\rrn_p$ can thus be written as
 \beq{phiMscPair}
 \phiMsc(\rrn_p|\calS) = {\sum_{\rrn_s\in\calS}}' \phiSc(\rrn_p|\rrn_s) + \calO(\tilde{\epsilon}^6).
 \eeq
Here, the primed summation means that the sum is performed in all
suspended particles except $s=p$, and the expansion parameter is
$\tilde{\epsilon} = \max_s\big\{\tilde{\epsilon}\big\}$.

\subsection{The acoustic interaction force}

When the particle interaction is taken into account through the scattered waves,
the radiation force can be written as the sum of contributions from
the unperturbed external field $\phiExt(\rrn_p)$ and from the configuration-dependent
interaction field, which involves terms like $\phiSExt(\rrn_p)\phiMsc(\rrn_p|\calS)$.
By substituting \eqref{phiMscS} into \eqref{FradOrder1} we find
 \bsub
 \beq{FradExtInt}
 \FFFrad(\rrn_p) =  \FFFradExt(\rrn_p) + \FFFradInt(\rrn_p|\calS).
 \eeq
The radiation force $\FFFradExt(\rrn_p)$ from the external field corresponds to $\phiIn =\phiExt$ in \eqref{FradOrder1},
\bal
  \nonumber
 \FFFradExt(\rrn_p) &=
 -\epsp^3 \pi \rhoO \nablabfTi_p
  \bigg[\!\frac{f_{0,p}}{3} \left|\phiExt(\rrn_p)\right|^2 \\
 &\quad
 -\frac{f_{1,p}}{2} \left|\nablabfTi_p \phiExt(\rrn_p) \right|^2 \! \bigg]
 + \calO\left(\tilde{\epsp}^5\right).
 \eal
It follows  that the configuration-dependent acoustic interaction force can be
expressed as a gradient force,
 \beq{FradInt}
 \FFFradInt(\rrn_p|\calS) = -\nablabf_p \: {\sum_ {\calS}}' U(\rrn_p|\rrn_s)
 + \calO\left(\tilde{\epsp}^5\right).
 \eeq
 \esub
For given probe and source positions $\rrn_p$ and $\rrn_s$
the pair-interaction potential is
\begin{align}
\nonumber
 U(\rrn_p|\rrn_s) &= \frac{\pi \epsp^3\: \rho_0}{k}
 \re\Big[  \frac{2 f_{0,p} }{3}
 \phiSExt(\rrn_p) \phiSc(\rrn_p|\rrn_s) \\
 &\quad
 - f_{1,p}\nablabfTi_p\phiSExt(\rrn_p) \cdot \nablabfTi_p\phiSc(\rrn_p|\rrn_s) \Big].
 \label{int_pot}
 \end{align}
%
%
The potential depends on a particle volume product
$a_p^3 a_s^3$, and on scattering factors like
$f_{i,p}f_{i,s}$ with $i=0,1$.
It is clear that the acoustic interaction force has the same dependence on these parameters.
Note further that the potential $U(\rrn_p|\rrn_s)$ is not necessarily
symmetric with respect to its indices.
Thus, the acoustic interaction force may not be symmetric either.

We now move on to analyze to which order in $\epsp$ the acoustic interaction force contributes to the total radiation force.
To ensure consistent approximations, this contribution must appear with a smaller order in $\epsp$ than the quadrupole
$\tilde{\epsp}^5$-contribution given in \eqref{FradOrder1}.
The pair-interaction approximation is more dominant when the dimensionless probe-source distance
$x_{ps}$ is small, satisfying $k(a_p+a_s)\le x_{ps}<1$.
Combining Eq.~(\ref{int_pot}) and \eqref{FradInt}, one
can show that the leading contribution to the interaction force is
\begin{equation}
\left|\FFFradInt\right|\sim
\left|\left[\nablabfTi_p\phiSExt(\rrn_p) \cdot \nablabfTi_p\right]\nablabfTi_p\phiSc(\rrn_p|\rrn_s)\right|
\sim
\frac{\epsp^3}{ x_{ps}^4}.
\end{equation}
We may express $x_{ps}$ in terms of the scattering parameter of the probe particle as
$x_{ps} = \gamma \epsp$, where $\gamma>1+a_s/a_p$.
Therefore, $|\FFFradInt| = \calO[\gamma^{-4} \epsp^{-1}]$.
Comparing the leading term in the acoustic interaction force with the quadrupole correction in \eqref{FradOrder1},
we find that consistent approximations are obtained, given that $\gamma^{-4}\epsp^{-1} \gg \tilde{\epsp}^5$
or that $\gamma$ is restricted to the limited range $1+a_s/a_p < \gamma \ll \epsp^{-3}$.
For example, if $\epsp=0.1$ then $\gamma \ll 1000$, otherwise the acoustic interaction force magnitude
becomes comparable to the quadrupole correction, which was already neglected in the radiation force expression
given in Eq.~(\ref{Frad}).

\section{Examples of the\\ acoustic pair-interaction force}
The acoustic interaction force exerted on a probe by a single source particle will  be determined considering
the interaction potential  in Eq.~(\ref{int_pot})
for an external plane traveling and standing wave.
The source particle is at the origin of the coordinate system $\vec{r}_s=\vec{0}$,
while the probe particle is at any other position $\vec{r}_p=\vec{r} = r \een_r$.
Furthermore, the shorthand notation
$U(\vec{r}) = U(\vec{r}|\vec{0})$
and $\FFFradInt(\vec{r}) = \FFFradInt(\vec{r}|\vec{0})$
will be used.

\subsection{Traveling plane wave}
Consider an external plane wave propagating along the $z$-axis.
The velocity  potential amplitude  of this wave is
 \begin{equation}
 \phi_\text{ext}(z) = \frac{v_0}{k} \ee^{\ii  k z },
 \label{phi_pw}
 \end{equation}
where
$v_0$ is  the magnitude of the particle velocity.

The  pair-interaction potential is calculated by substituting
Eq.~(\ref{phi_pw}) into Eq.~(\ref{phi_scatt}).
Thus, inserting the obtained result into Eq.~(\ref{int_pot}),
we find in spherical coordinates $(r,\theta,\varphi)$ that
\begin{widetext}
\begin{align}
\nonumber
U(r, \theta) &= \frac{\pi E_0k^2 a_p^3a_s^3}{ r}\biggl[ \cos\big[k r (1-\cos \theta)\big]
\left(\frac{ 3f_{1,p} f_{1,s} (1-3\cos^2\theta) + f_{0,p} f_{1,s} \cos \theta}{  3 k r}
+\frac{6f_{1,p} f_{0,s} \cos \theta - 2 f_{0,p} f_{0,s}}{9}\right)\\
& \quad
+  \sin \big[k r (1-\cos \theta )\big] \left(\frac{ f_{1,p} f_{1,s}(3 \cos^2\theta-1)  }{(k r)^2}
-\frac{2 f_{1,p} f_{0,s} \cos \theta  }{3 k r} -f_{1,p} f_{1,s} \cos^2\theta
+\frac{f_{0,p} f_{1,s} \cos \theta  }{3 }\right)\biggr],
\label{Uint_pw}
\end{align}
\end{widetext}
where $E_0=\rho_0 v_0^2/2$ is the characteristic energy density
of the external traveling plane wave. Below we study two special cases of this expression, and in this context it is useful to introduce the compression and density interaction potential strengths, $\UpsO$ and $\UpsI$, respectively,
 \bsubal
 \eqlab{UpsOdef}
 \UpsO &= \frac{2\pi}{9}\: E_0  k^3 a_p^3a_s^3\: f_{0,p}f_{0,s},\\
 \eqlab{UpsIdef}
 \UpsI &= \pi\: E_0  k^3 a_p^3a_s^3\: f_{1,p}f_{1,s}.
 \esubal

As the first special case, we reproduce the seminal result for the secondary Bjerknes force between two bubbles, for which it is assumed that the external wave frequency is much smaller than the
resonance frequency of the bubbles. Since for gas bubbles $f_0 \approx -10^5$ and $f_1 \approx -2$, and because $kr > ka_s \approx 10^{-3}$ implies that $f_0 \gg f_1/(kr)$, only the term involving $f_{0,p}f_{0,s}$ is relevant in Eq.~(\ref{Uint_pw}), and we arrive at
at $U$ and $\FFFradInt = -\nablabf U$,%
 \bsub
 \bal
 &\qquad \qquad U(r,\theta) = -\UpsO \:
 \frac{\cos\big[kr(1-\cos\theta)\big]}{kr},\\
 &\FFFradInt(r,\theta) = -k \UpsO \Bigg[
 \frac{\sin\big[kr(1-\cos\theta)\big]\sin\theta}{kr}\:
 \een_\theta\\
 \nonumber
 &+ \frac{\cos\!\big[kr(1\!-\!\cos\theta)\big] \!+\!
 kr\sin\!\big[kr(1\!-\!\cos\theta)\big](1\!-\!\cos\theta)}{(kr)^2}\een_r\Bigg]
 \\
 &\approx -\frac{k\UpsO}{(kr)^2}\eee_r =
 -\frac{2 \pi E_0  k^2 a_p^3a_s^3 \kappa_{p}\kappa_{s}}{9 \kappa_0^2 r^2 }\eee_r,
 \quad k r \ll 1,
 \eal
 \esub
where the latter is the secondary Bjerknes force in the short-range limit as derived by Zheng and Apfel \cite{Zheng1995}.

As the second special case, we consider the acoustic interaction between particles collected, say, by the primary acoustic force, in the transverse plane $(\theta=\pi/2)$. Since the phase of the external wave does not change in the transverse plane, the angular dependence drops out of  Eq.~(\ref{Uint_pw}) in this special case, and only the radial distance $\varrho=\sqrt{x^2+y^2}$ in the transverse plane and the associated in-plane radial unit vector $\eee_\varrho$ play a role in the following. The potential $U$ becomes
 \bsub
 \begin{align}
 \label{Uint_pw_transverse}
 U(\varrho) &= \UpsO\: n_0 (k \varrho)
 - \UpsI\: \frac{j_1(k \varrho)}{k\varrho},
 \end{align}
where $n_0(x) = -\cos(x)/x$ is the zero-order spherical Neumann function and $j_1(x) = \sin(x)/x^2 - \cos(x)/x$ is the first-order spherical Bessel function. In the short range limit $k \varrho \ll 1$, minus the gradient of Eq.~(\ref{Uint_pw_transverse}) gives
 \begin{equation}
 \FFFradInt(\varrho) = -k\UpsO\: \bigg[\frac{1}{ (k\varrho)^2}
 + \calO (1)\bigg]\eee_\varrho, \quad k\varrho \ll 1,
 \label{Upw_zero}
 \end{equation}
which depends quadratically on both the inverse inter-particle distance and on the frequency, and which is antisymmetric $\FFFradInt(\vec{r}_p|\vec{r}_s) =
-\FFFradInt(\vec{r}_s|\vec{r}_p)$. In the long-range linit $k \varrho \gg 1$,
the acoustic interaction force in the transverse plane is
 \begin{align}
 \FFFradInt(\varrho) &=
 -k \UpsO\: \bigg[\frac{\sin(k \varrho)}{k \varrho}
 + \calO \big([k\varrho]^{-2}\big)\bigg], \quad k\varrho \gg 1.
 \end{align}
 \esub
This result has been previous obtained by Zhuk for the interaction of two rigid particles
($f_{0,s}=f_{0,p}=1$)~\cite{Zhuk1985}. Note that the acoustic interaction force decays with the inter-particle distance, but that it oscillates in space with two consecutive zeros separated by a half wavelength of the external plane traveling wave. It should be noticed that the only mechanical property that affects the acoustic interaction force on both short-range and  long-range limits is the  compressibility of the particles.

\subsection{Standing plane wave}
\label{sec:sw}
Now, consider the case, where the external incident wave is a standing plane wave defined by the potential
\begin{equation}
\label{phi_sw}
\phi_\text{ext}(z) =  \frac{v_0}{k} \sin \big[k (z-h) \big],
\end{equation}
where $h$ is the distance from the first wave node to the origin of the coordinate system. A particle  exposed to such a wave will be collected in the potential node if the scattering coefficients satisfy $ 2f_{0,p} < - 3f_{1,p} $ and in the potential antinode if $2f_{0,p} > - 3f_{1,p}$.

We calculate the interaction potential  by inserting Eqs.~(\ref{phi_scatt}) and~(\ref{phi_sw})
into Eq.~(\ref{int_pot}).
Accordingly, we obtain
\begin{widetext}
\begin{align}
\nonumber
 U(r, \theta) &= \frac{\pi E_0 k^2a_p^3a_s^3}{ r}\\
\nonumber
 &\times \biggl\{
\sin [k (r \cos \theta -h)]
\biggl[2 f_{0,p} f_{1,s} \cos (k h) \cos \theta \frac{ \sin (k r)  }{3 k r} +
\left(\frac{4}{9} f_{0,p} f_{0,s} \sin (k h)
-\frac{2}{3} f_{0,p} f_{1,s} \cos (k h) \cos \theta  \right) \cos (k r) \biggr]\\
\nonumber
&+
\cos [k (r \cos \theta -h)]
\biggl[ f_{1,p} f_{1,s} \cos (k h)(3 \cos^2\theta - 1 ) \frac{\sin (k r)}{(k r)^2}
+ \biggl(\frac{2}{3} f_{1,p} f_{0,s} \sin (k h) \cos \theta + f_{1,p} f_{1,s} \cos (k h)\\
&-3 f_{1,p} f_{1,s}  \cos (k h) \cos^2\theta  \biggr)\frac{\cos k r}{k r}
+\left[\frac{2}{3} f_{1,p} f_{0,s} \sin (k h)
-f_{1,p} f_{1,s} \cos (kh) \cos \theta \right] \cos \theta \frac{\sin (k r) }{k r}\biggr]
\biggr\}.
\label{Uint_sw}
\end{align}
\end{widetext}

As above and using a similar analysis, we first study the acoustic interaction force between two air bubbles. The force is given by minus the gradient of  Eq.~(\ref{Uint_sw}) considering only the term containing $f_{0,p}f_{0,s}$, and we arrive at the secondary Bjerknes force in a standing plane wave,
 \beq{FBjerknessSW}
 \FFFradInt(r) \approx -\frac{2 \pi  E_0 k^2 a_p^3a_s^3 \kappa_{p}\kappa_{s}}{
 9 \kappa_0^2 r^2 }\sin^2(k h)\eee_r, \; k r\ll 1.
\eeq
This is equivalent to the result obtained by
Zheng and Apfel \cite{Zheng1995}.

Next, we focus on the acoustic interaction force between particles in the transverse plane defined by $\theta = \pi/2$. In this special case, Eq.~(\ref{Uint_sw}) reduces to
\begin{align}
U\left(r,\frac{\pi}{2}\right) &= 2\UpsO\:\sin^2(k h) n_0 (k r)
- \UpsI\:\cos ^2(k h)\frac{j_1(k r)}{kr}.
\label{Uxy_sw}
\end{align}
We note that this interaction potential only depends on distance between the source and the probe, and consequently, the acoustic interaction force between particles in the transverse plane is
antisymmetric with respect to the probe and the source particles.

According to whether a given set of particles are collected in either the nodal or the antinodal planes of the standing wave, we can choose to let the transverse plane coincide with a nodal plane by setting $kh=0$, in which case all $\sin(kh)$-terms vanish in Eq.~(\ref{Uxy_sw}), and with an antinodal plane by $kh=\pi/2$, in which case all $\cos(kh)$-terms vanish.

Thus, from the gradient of $U$ in Eq.~(\ref{Uxy_sw}) we obtain the acoustic interaction force between particles in the nodal plane ($kh = 0$) in the short-range limit $k\varrho \ll 1$ to be
 \bsub
 \begin{equation}
 \FFFradInt(\varrho) = -\frac{1}{15}\:k \UpsI\: \Big[ k \varrho  + \calO\big([k\varrho]^3\big)\Big]\:\eee_\varrho,
 \end{equation}
which has a strong fifth-power frequency dependence and a linear dependence on the inter-particle distance. Only the density scattering factors and not the compressibility factors enters. In the long-range limit $k\varrho \gg 1$ for the nodal plane, the acoustic interaction force is
 \begin{equation}
 \FFFradInt(\varrho) = k\UpsI\: \Bigg[\frac{\sin(k \varrho)}{(k\varrho)^2}
     +\calO\big([k\varrho]^{-4}\big) \Bigg] \eee_\varrho,
 \end{equation}
 \esub
which has an oscillatory behavior with half an external wavelength distance between two consecutive zeros,
while it decays with the inverse-square of the inter-particle distance. It depends quadratically on the  frequency and only the density scattering factors, and not the compressibility factors, appear.

Similarly, in the antinodal plane ($kh = \pi/2$), the short-range limit $k\varrho \ll 1$ of the acoustic interaction force is
 \bsub
 \begin{equation}
 \FFFradInt(\varrho) = -2k\UpsO\:
 \bigg[\frac{1}{(k \varrho)^2} + \calO(1) \bigg] \eee_\varrho,
 \end{equation}
while the long-distance limit $k\varrho \gg 1$ is
 \begin{equation}
 \FFFradInt(\varrho) = -2k\UpsO\: \frac{\sin(k \varrho)}{k\varrho}
 \eee_\varrho.
 \label{RF_antinodal}
 \end{equation}
 \esub
We note that in the antinodal plane only the compressibility scattering factors appear.

\begin{figure*}
\centering
\includegraphics[]{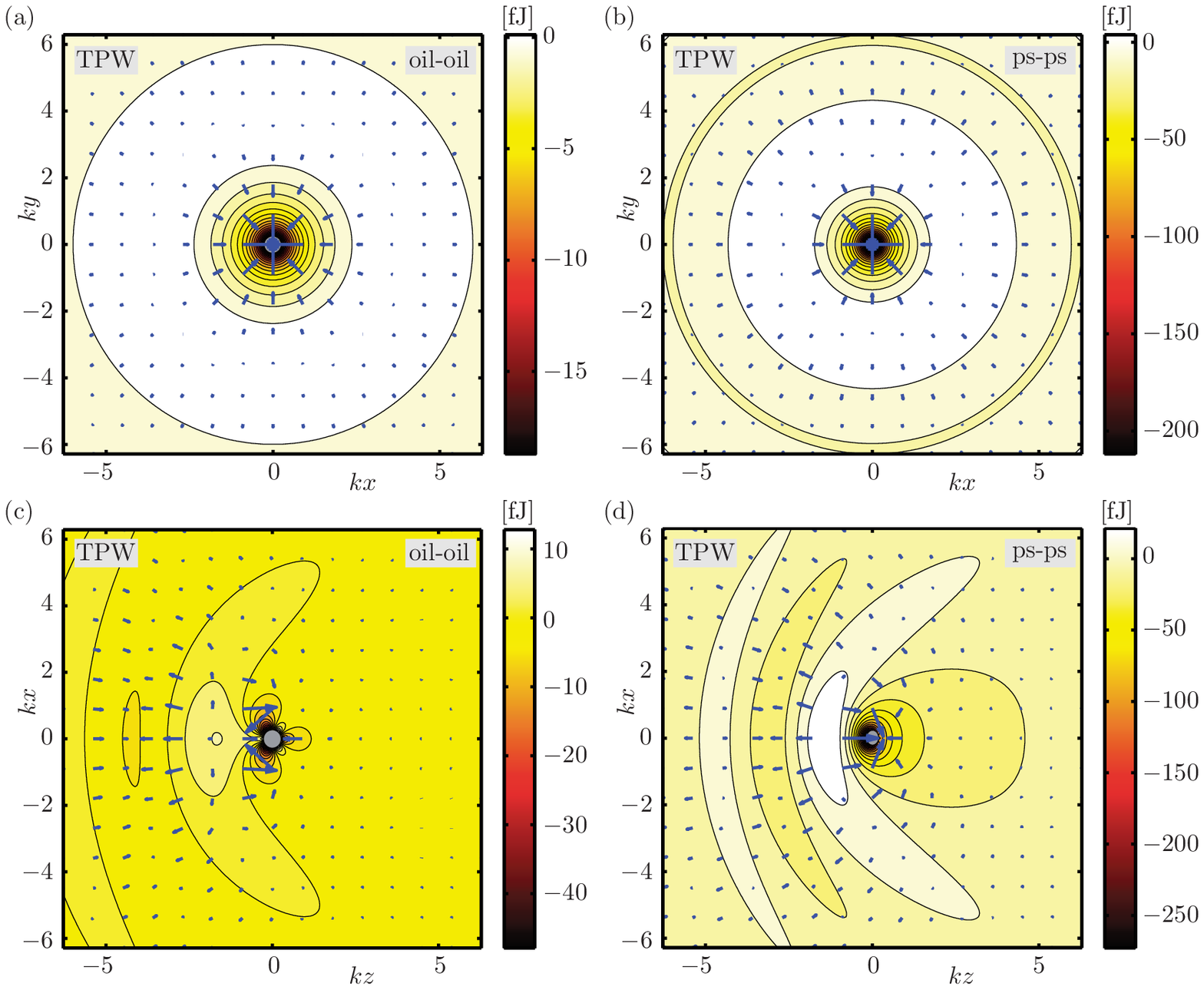}
\caption{(Color online) The acoustic interaction pair potential $U(\rrn_p|\zerovec)$ [Eq.~(\ref{Uint_pw}), contours] and force $\FFFradInt(\rrn_p|\zerovec) = -\nablabf U$ [arrows] between a pair of identical 12-$\SImum$ particles induced by the traveling plane wave (TPW) Eq.~(\ref{phi_pw}), with the source particle located at the origin, $\rrr_\text{s}=\vec{0}$, for $kr > 0.2$. (a) Silicone oil droplets (oil-oil), with the probe $\rrn_p = (x,y,0)$ in the transverse $xy$-plane. (b) Same as (a) but for polystyrene microparticles (ps-ps). (c) Same as (a), but with the probe $\rrn_p = (x,0,z)$ in the parallel $xz$-plane (oil-oil). (d) Same as (b), but with the probe $\rrn_p = (x,0,z)$ in the parallel $xz$-plane (ps-ps).}
\label{fig:pw_result}
\end{figure*}

\section{Mean-field approximation}
Going beyond the simple two-particle problem, we now derive an analytical expression for the acoustic interaction force between a probe particle and the particles surrounding it in a homogeneous particle suspension. In Section~\ref{sec:ScatteringSuspension} we considered $N$ particles with positions $\rrn_s$ in a given configuration $\calS$ in a suspension of volume $V$. Using Dirac's delta function $\delta(\rrr)$, we can formally rewrite the sum $\calU$ over pair-potentials $U$ as an integral,
 \bsubal
 \label{Uint}
 \calU(\rrn_p) &= {\sum_{\rrn_s\in\calS}\!}' U(\rrn_p|\rrn_s) =
 \int_V U(\rrn_p|\rrr)\: n(\rrr)\;\dm \rrr,
 \\ \label{Umean}
 n(\rrr) &= {\sum_{\rrn_s\in\calS}\!}' \delta(\rrr-\rrn_s),
 \esubal
where $n(\rrr)$ can be interpreted as the particle concentration field. In a mean-field approximation, $n(\bm{r})$ is smoothened, such that the number $\dm N$ of particles particles in a small volume $\dm \rrr $ at any given position $\bm{r}$ is given by $\dm N = n(\bm{r})\: \dm \rrr $. For a homogeneous suspension, we have $n(\rrr) \approx N/V$, and the interaction potential $\calU(\rrn_p)$ experienced by  the probe particle is well approximated by
 \begin{equation}
 \calU(\rrn_p) \approx
 \frac{N}{V} \int_V U(\rrn_p|\rrr) \;\dm \rrr.
\end{equation}
This mean-field approximation is expected to improve for an increasing number of source particles per volume.

To illustrate the mean-field approximation in the acoustic interaction force problem,
we assume that the source particles are uniformly distributed within a circular region of radius $R$ and thickness $2a_s$ at the antinodal plane (the $xy$-plane) of the external standing plane wave Eq.~(\ref{phi_sw}). The volume occupied by the particle distribution is thus $V=2\pi R^2 a_s$. The probe particle is placed at the origin of the coordinate system, while the center of the disk-shaped source-particle region is displaced backwards along what is defined to be the $x$-axis to the position $-r_p\:\een_x$. With this configuration and using  Eq.~(\ref{Uxy_sw}), the pair-interaction potential $U(\rrn_p|\rrn_s)$ becomes
 \begin{equation}
 U(\rrn_p|\rrn_s) = -2\UpsO\:\frac{\cos k r_s}{kr_s}, \text{ with } \rrn_p = \zerovec,
 \label{U_ps}
 \end{equation}
while the limits of the integration region $V$ in the expression~(\ref{Umean}) for the total interaction potential requires some analysis. Using the cylindrical polar coordinates $(r,\varphi,z)$ for the source position $\rrn_s$, we find that in the direction $\varphi$, a source particle can at most be at the distance $R'(\varphi)$ from the probe particle,
 \beq{Rmax}
 R'(\varphi) = \sqrt{R^2 - x^2_p\sin^2\varphi} - r_p\cos\varphi.
 \eeq
The total interaction potential $\calU$ having a strength of $\calU_0 = 2N\UpsO/\pi$ for the probe particle at $\rrn_p=\zerovec$, becomes
 \bal
 \nonumber
 \calU(\rrn_p) &= -\frac{2 N \UpsO}{\pi R^2 (2a_s)}\:
 \int_0^{2\pi}\!\dm \varphi \int_{-a_s}^{a_s}\!\dm z \int_0^{R'(\varphi)} \!\dm r\:
 r\:\frac{\cos(kr)}{kr}  \\
 \label{UmeanDisk}
 &= -\calU_0
 \int_0^{2\pi}\!\dm \varphi\: \sin\big[kR'(\varphi)\big], \quad
 \calU_0 = \UpsO\:\frac{2N}{\pi}.
 \eal
For an arbitrary position $\rrn_p$ of the probe particle relative to the center of the source-particle region, this integral can be evaluated numerically. However, for small displacements $r_p \ll R$, we can obtain an analytical expression by Taylor expanding the integrand,
 \bal
 \sin\big[kR'(\varphi)\big] &\approx
 \sin kR - kr_p \cos\varphi \cos kR
 \\ \nonumber
 &\quad - \frac{\sin^2\varphi\cos kR + kR\cos^2\varphi\sin kR}{2kR}\:(kr_p)^2,
 \eal
which upon insertion into Eq.~(\ref{UmeanDisk}) leads to
 \bal
 \nonumber
 \calU(\rrn_p) = &-\calU_0
 \bigg[\frac{\sin kR}{(kR)^2} \!-\! \bigg(
 \frac{\sin kR}{(kR)^2} \!+\!  \frac{\cos kR}{(kR)^3}\bigg)\frac{(kr_p)^2}{4}\bigg] \\
 &+ \calO\big[(kr_p)^4\big].
  \label{UmeanAprrox}
 \eal
By taking minus the gradient $-\nablabf_p = -\eee_r\partial/\partial_{r_p}$ relative to the probe position,
we determine the acoustic interaction force on the probe particle to be
 \beq{rf_meanfield_r_zero}
 \FFF_\mathrm{rad}^\mathrm{int}(\rrn_p) =-\frac{k\calU_0}{2}
  \bigg[ \frac{\sin kR}{(kR)^2} +  \frac{\cos kR}{(kR)^3}\bigg]k\rrn_p
 + \calO\big[(kr_p)^3\big].
 \eeq
We note that, as expected, the interaction force is zero in the case $r_p = 0$, where the source particles are symmetrically distributed around the probe particle. Moreover, the interaction force tends to zero in the limit $kR \gg 1$ for fixed $kr_p$, a fact that can be explained by the decreasing degree of asymmetry in the source particles characterized by the decreasing ratio $r_p/R$.
We also note that if the sign of the compressibility factors $f_{0,p}$ and $f_{0,s}$ are the same, the symmetric position $r_p = 0$ is a stable equilibrium point if $\cos kR + k R \sin kR >0$, in which case the particles will be attracted to the center of the source region. Finally, we note that the frequency dependence of the interaction force is governed by the trigonometric factors. In the case of a small disk region, $kR \ll 1$, we have that $(\cos kR)/(kR)^3 \approx (kR)^{-3}$ dominates. Consequently, in this case the acoustic interaction force depends on the wavenumber as $k^2$ and thus quadratically with frequency.

\begin{figure*}
\centering
\includegraphics[]{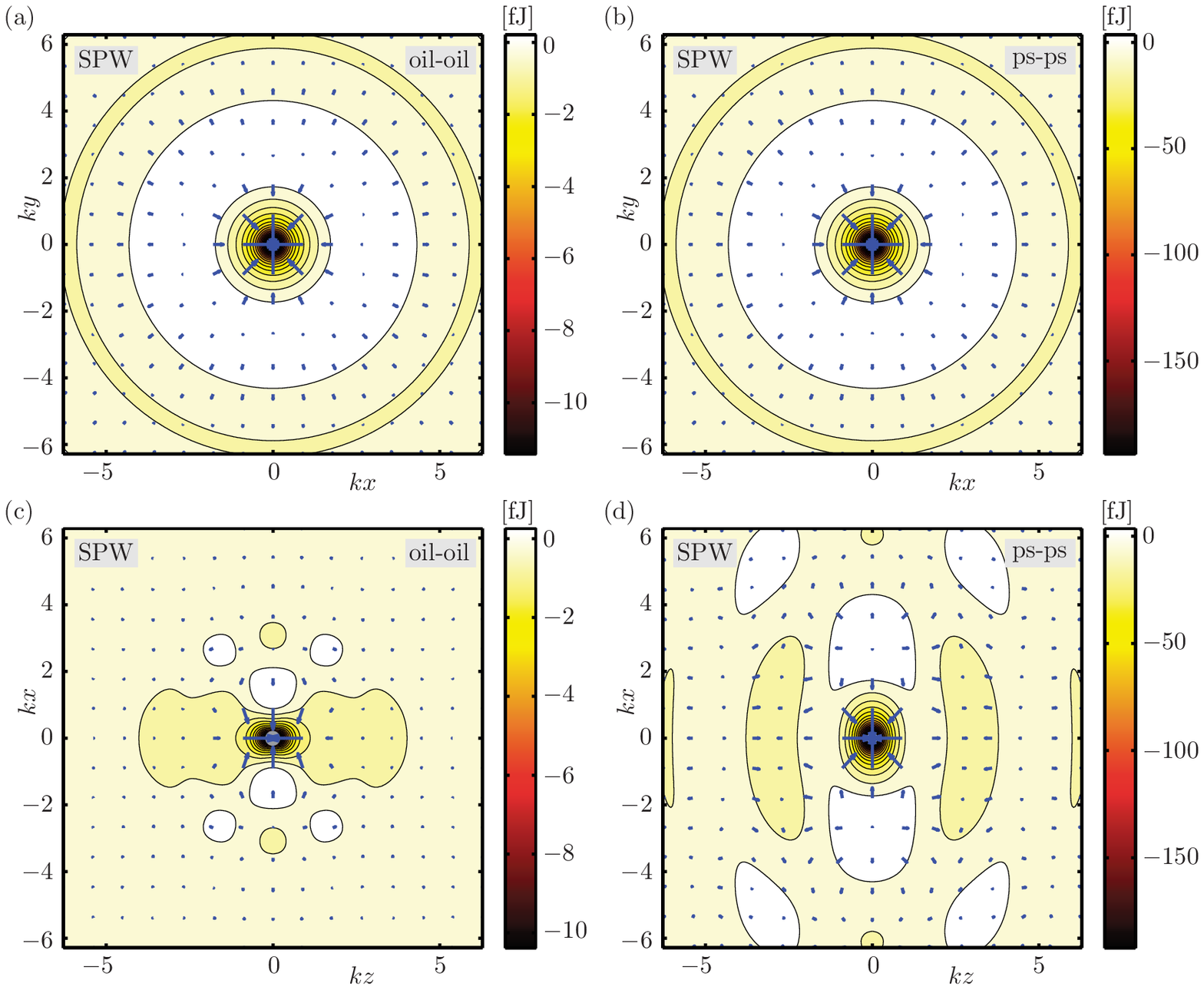}
\caption{(Color online) The acoustic interaction pair potential $U(\rrn_p|\zerovec)$ [Eq.~(\ref{Uint_sw}), contours] and force $\FFFradInt(\rrn_p|\zerovec) = -\nablabf U$ [arrows] between a pair of identical 12-$\SImum$ particles induced by the standing plane wave (SPW) Eq.~(\ref{phi_sw}) with $kh=\pi/2$ and the source particle located at the origin, $\rrr_\text{s}=\vec{0}$ for $kr>0.2$. (a) Silicone oil droplets (oil-oil), with the probe $\rrn_p = (x,y,0)$ in the transverse $xy$-plane. (b) Same as (a) but for polystyrene microparticles (ps-ps). (c) Same as (a), but with the probe $\rrn_p = (x,0,z)$ in the parallel $xz$-plane (oil-oil). (d) Same as (b), but with the probe $\rrn_p = (x,0,z)$ in the parallel $xz$-plane (ps-ps).}
\label{fig:sw_result}
\end{figure*}

\section{Results and discussion}
In this section, based on direct numerical evaluations of $U(\rrr|\zerovec)$ in Eqs.~(\ref{Uint_pw}) and~(\ref{Uint_sw}) for a  traveling and standing place wave, respectively, we calculate  the acoustic interaction force between a pair of silicone oil droplets and a pair of polystyrene microparticles suspended in water at room temperature. The water is characterized by its density $\rho_0=\unit[1000]{kg/m^3}$ and speed of sound $c_0=\unit[1500]{m/s}$.
Using the material parameters of Ref.~\cite{kino1987}, the scattering factors $f_0$ and $f_1$ defined in Eq.~(\ref{factors})
are found to be $(f_0,f_1) = (-0.08, 0.07)$ for silicone oil and $(f_0,f_1) = (0.46, 0.038)$ for polystyrene.
For the external wave, we choose the following typical parameter values from actual acoustophoresis
experiments~\cite{Barnkob2010}:
frequency  $\omega/(2 \pi) = \unit[2]{MHz}$, wavenumber $k = 8378~\SIm^{-1}$,
and energy density $E_0=\unit[10]{J/m^3}$.
The microparticle radius is $a_p = a_s = \unit[12]{\SImum}$, so we obtain $ka_s = ka_p=0.1$.
Below, the source particle is positioned at $\rrr_\textbf{s} = \vec{0}$,
whereas the probe is placed at any position in space, $\rrr_p = \rrr$.

We compute the acoustic interaction force $\FFFradInt$ due to an external traveling wave or an standing wave plane, from the pair potential $U(\rrr|\zerovec)$ in Eqs.~(\ref{Uint_pw}) and (\ref{Uint_sw}), respectively, as $\FFFradInt = -\nablabf U(\rrr|\zerovec)$  using Mathematica software~\cite{Mathematica}. The probe position $\rrn_p$ is presented in the scaled Cartesian coordinates $k\rrn_p = (kx, ky, kz)$.

\subsection{Particle pairs in a traveling plane wave}
In Fig.~\ref{fig:pw_result}, we show the pair potential $U(\rrr|\zerovec)$ (contour plot) and the corresponding radiation force $\FFFradInt$ (arrows) induced by the external traveling plane wave Eq.~(\ref{phi_pw})
propagating along the $z$-axis for a pair of oil microdroplets and a pair of polystyrene microparticles, respectively. In the (transverse) $xy$-plane Fig.~\ref{fig:pw_result}(a) and (b), the acoustic interaction force is central and also  attractive in the region $kr<2$ for both situations. The force is central because the microparticles directly interact through their scattered waves. Note that in the short-range distance, the acoustic interaction force is about $-U/a_p$. Hence, the force magnitude on the oil and the polystyrene probe microparticles is less than $\unit[1.6]{nN}$ and $\unit[18]{nN}$, respectively. In the (parallel) $xz$-plane Fig.~\ref{fig:pw_result}(c) and (d), the situation is different. The acoustic interaction force is not central force, because the scattered waves interact with the external waves whose phase varies along the $z$-axis. Most of the potential variation occurs in the backscattering direction of the source microparticle $kz<0$. This happens because in the Rayleigh
scattering most of the incident plane traveling wave is backscattered~\cite{Pierce1989}. Note that the potential forms attractive islands for the probe microparticles at $kz=-2.5$ and $kz\sim 0$. The magnitude of $\FFFradInt$ between the oil and the polystyrene probe microparticles is less than $\unit[4.1]{nN}$ and $\unit[23]{nN}$, respectively.

\begin{figure}
\centering
\includegraphics[]{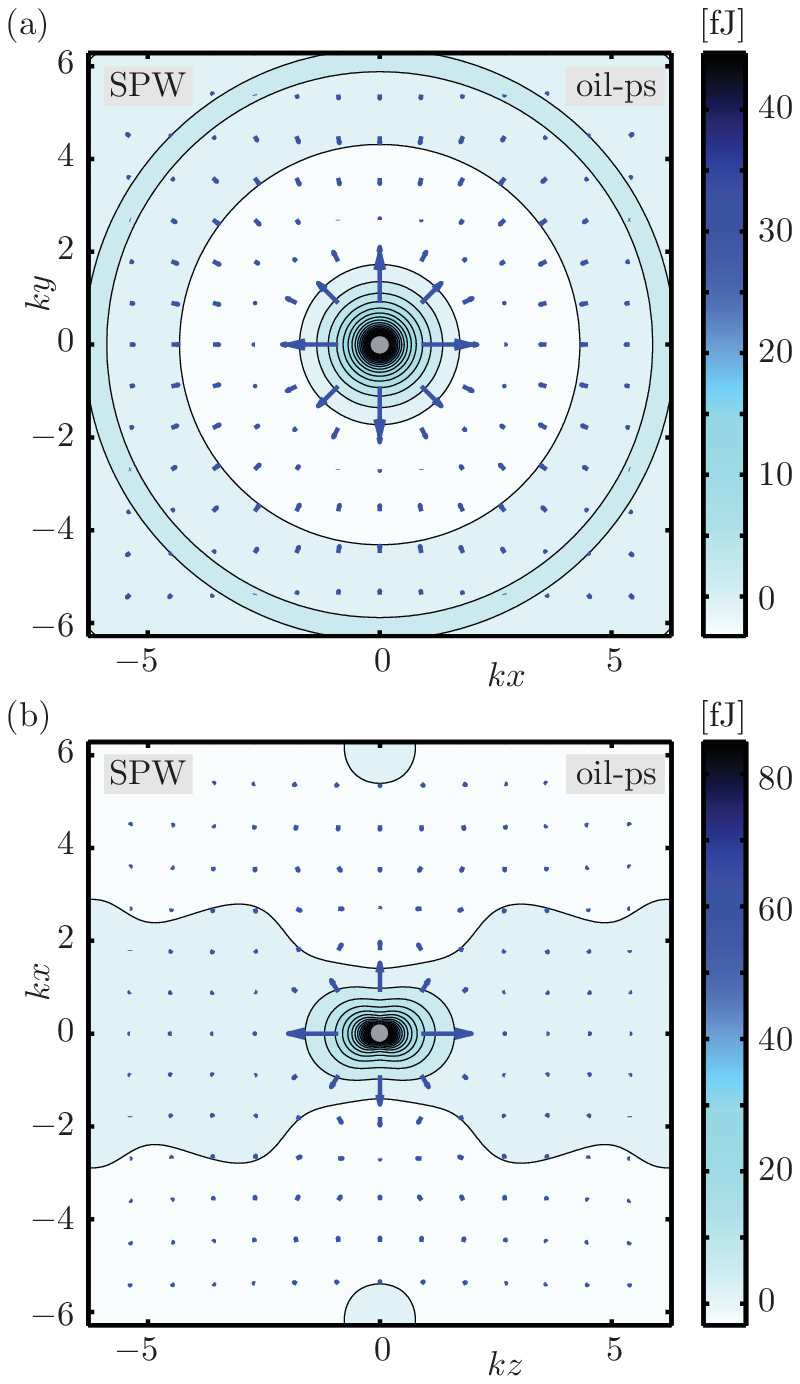}
\caption{(Color online) Same as Fig.~\ref{fig:sw_result}(a) and (c), but with two different particles (oil-ps): a 12-$\SImum$ silicone oil droplet as the probe particle at $\rrn_p = \rrr$ and a 12-$\SImum$ polystyrene particle as the source particle at $\rrr = \zerovec$.}
\label{fig:sw_ps_so_result}
\end{figure}

\subsection{Particle pairs in a standing plane wave}
In Fig.~\ref{fig:sw_result},  we show the pair potential $U(\rrr|\zerovec)$ (contour plot) and the corresponding radiation force $\FFFradInt$ (arrows) induced by the external standing plane wave Eq.~(\ref{phi_sw}) along the $z$-axis with $k h = \pi/2$ resulting in an antinode in the transverse $xy$-plane for a pair of oil microdroplets and a pair of polystyrene microparticles, respectively. For both particle pairs, the primary radiation force focus particles in the antinodal plane as discussed in Sec.~\ref{sec:sw}
In the transverse $xy$-plane Fig.~\ref{fig:sw_result}(a) and (b), the acoustic interaction force is attractive when $kr<5$ for the oil microdroplets and when $kr<2$ for the polystyrene microparticles. In the parallel plane, energy potential wells are formed around $kz=0$ and $\pm3.6$ for the polystyrene particles. In these regions, the probe microparticles might be trapped. In both situations, the interaction is attractive in the vicinity of the source microparticle.  The magnitude of $\FFFradInt$ between the oil and the polystyrene probe microparticles is less than $\unit[1.0]{nN}$ and $\unit[17]{nN}$, respectively.

For the same standing plane wave, we present in Fig.~\ref{fig:sw_ps_so_result} the acoustic interaction force between two different particles, a polystyrene source  and a silicone oil probe. In the transverse $xy$-plane Fig.~\ref{fig:sw_ps_so_result}(a), the acoustic interaction force is repulsive in a region $kr<2$, while outside this region, the force becomes mostly attractive. The radiation force magnitude is less than $\unit[5]{nN}$. In the parallel $xz$-plane Fig.~\ref{fig:sw_ps_so_result}(b), the main role of the acoustic interaction force is to be repulsive in the region $kx<1$ and $kz<1.5$.

\begin{figure}[b!]
\centering
  \includegraphics[]{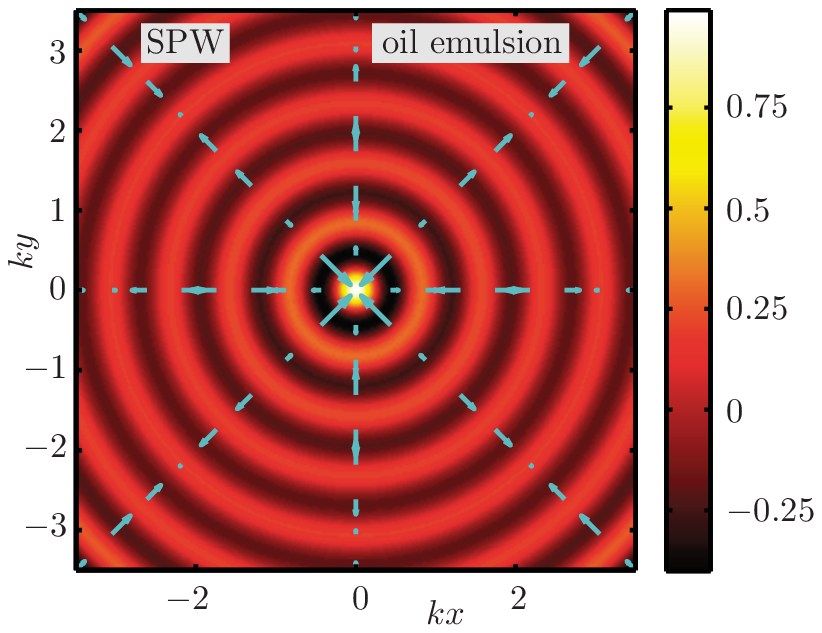}
  \caption{(Color online) The normalized interaction potential  $\calU/\calU_0$ (countour plot)
  of the acoustic interaction force $\FFFradInt$ (arrows) in an aqueous emulsion of soybean oil droplets of
  radius $a_s=\unit[12]{\SImum}$ in the external standing plane wave (SPW) defined in Eq.~(\ref{phi_sw}).
  The emulsion consists of $N \approx 3800$ droplets in a cylindrical disk region of radius $R=\unit[5]{mm}$
  and thickness $2a_s$ at the nodal $xy$-plane. Here, $\calU_0 = \unit[0.2]{fJ}$.
  }
  \label{fig:rf_oil_water}
\end{figure}

\subsection{Emulsion of oil droplets in water}

As a last numerical example, we consider the multi-particle system comprising an emulsion of soybean oil
droplets of radius $a_s=\unit[12]{\SImum}$ in water.
The scattering factors are $f_{0,s}=-0.11$ and $f_{1,s}=-0.06$ \cite{coupland1997},
so according to Sec.~\ref{sec:sw}, the oil microdroplets will collect in a node  when exposed to a standing plane wave.
We therefore study the effects of such a wave described by Eq.~(\ref{phi_sw}) with $kh = 0$ droplets
initially uniformly distributed  in the transverse (nodal) $xy$-plane in a circular disk-shaped region of radius
$R=\unit[5]{mm}$ and thickness $2a_s$.
The mean  interdroplet distance is assumed to be $10 a = \unit[120]{\SImum}$,
which corresponds to $N\approx3800$ droplets within the disk-shaped region.
The interaction potential is then calculated by numerical integration of the mean-field approximation~(\ref{UmeanDisk})
for $\calU$. Subsequently, the acoustic interaction force was determined by calculating numerically minus the gradient
of this $\calU$. In this case the pair interaction potential has the strength has the value
$\UpsO = \unit[8.1\times10^{-20}]{J}$ yielding a total interaction strength of $\calU_0= \unit[0.2]{fJ}$.
Hence, the magnitude of the acoustic interaction force in the short-range distance is about
$\calU_0/a_s = \unit[16]{pN}$.

In Fig.~\ref{fig:rf_oil_water}, we depict the normalized interaction potential  $\tilde{\calU}=\calU/\calU_0$ (contour plot) and the associated acoustic interaction force $\FFFradInt$ (arrows) on a probe droplet placed in the transverse $xy$-plane in the disk-shaped region of source droplets. The potential $\tilde{\calU}$ exhibits concentric local maxima and minima, with the global maximum  localized at $kr = 0$. Hence, microdroplets have the tendency to move away from the central region.
This is in agreement with \eqref{rf_meanfield_r_zero}, because here $(\cos kR + k R \sin kR)\approx -36$. On the other hand, the potential minima will attract the nearby oil microdroplets.
Therefore,  microdroplets may aggregate in the concentric regions of minima throughout the emulsion. Note that the distance between to consecutive minima is about $10\%$  of the incident wavelength.
Furthermore, the magnitude of the acoustic interaction force is less than $\unit[0.02]{nN}$.

\section{Summary and conclusion}
We have developed a potential theory for the acoustic interaction forces
in a collection of $N$ suspended particles in an ideal fluid, considering
the long-wavelength limit $a_s\ll \lambda$ $(s=1,2\dots,N)$.
The particles were considered to be  compressible fluid or elastic solid spheres.

In our analysis, the acoustic interaction force between two particles
is expressed in terms of minus the gradient of a pair-interaction potential.
In turn, this function depends on the product of the external and scattering velocity potentials.
We have shown that the multi-scattering contribution to the acoustic interaction force
on a particle placed at $\bm{r}_p$
is dominated by scattering waves having undergone only a single prior scattering event due
to a source particle located at $\bm{r}_s$ $(s\neq p)$.

The investigations of the interaction between particle pairs in a traveling or standing plane wave
have shown that the acoustic interaction forces might by  attractive or repulsive for  short-range
interaction. In the transverse $xy$-plane to the wave propagation direction, the acoustic force is a central force,
while in in the parallel $xz$-plane this does not happen. The short-range attraction or repulsive roles of the acoustic interaction force are determined by the compressibility of the particles.

To address the many-particle case, we have presented a mean-field theory based on the continuous limit of the acoustic interaction  potential. Analytical results have been obtained for a symmetric suspension of source particles with the probe placed near the center of the suspension. In general, numerical evaluation of the mean-field expression are necessary, and as an example of this, we studied an emulsion formed by oil droplets in water. Under a standing plane wave, oil droplets have the tendency to cluster in concentric regions on the transverse $xy$-plane.

The theoretical predictions discussed in this work might be confirmed in acoustophoresis experiments using ultrasonic demulsification techniques~\cite{nii2009} or by a combination of micro-particle image velocimetry \cite{Augustsson2011} and frequency tracking \cite{Hammarstrom2014}.

\acknowledgments
This work was supported by the Danish Council for Independent Research,
Technology and Production Sciences, Grant No.~11-10702 and by CAPES (Brazilian Agency), Grant No.~17997-12-7.


\begin{thebibliography}{35}%
\makeatletter
\providecommand \@ifxundefined [1]{%
 \@ifx{#1\undefined}
}%
\providecommand \@ifnum [1]{%
 \ifnum #1\expandafter \@firstoftwo
 \else \expandafter \@secondoftwo
 \fi
}%
\providecommand \@ifx [1]{%
 \ifx #1\expandafter \@firstoftwo
 \else \expandafter \@secondoftwo
 \fi
}%
\providecommand \natexlab [1]{#1}%
\providecommand \enquote  [1]{``#1''}%
\providecommand \bibnamefont  [1]{#1}%
\providecommand \bibfnamefont [1]{#1}%
\providecommand \citenamefont [1]{#1}%
\providecommand \href@noop [0]{\@secondoftwo}%
\providecommand \href [0]{\begingroup \@sanitize@url \@href}%
\providecommand \@href[1]{\@@startlink{#1}\@@href}%
\providecommand \@@href[1]{\endgroup#1\@@endlink}%
\providecommand \@sanitize@url [0]{\catcode `\\12\catcode `\$12\catcode
  `\&12\catcode `\#12\catcode `\^12\catcode `\_12\catcode `\%12\relax}%
\providecommand \@@startlink[1]{}%
\providecommand \@@endlink[0]{}%
\providecommand \url  [0]{\begingroup\@sanitize@url \@url }%
\providecommand \@url [1]{\endgroup\@href {#1}{\urlprefix }}%
\providecommand \urlprefix  [0]{URL }%
\providecommand \Eprint [0]{\href }%
\providecommand \doibase [0]{http://dx.doi.org/}%
\providecommand \selectlanguage [0]{\@gobble}%
\providecommand \bibinfo  [0]{\@secondoftwo}%
\providecommand \bibfield  [0]{\@secondoftwo}%
\providecommand \translation [1]{[#1]}%
\providecommand \BibitemOpen [0]{}%
\providecommand \bibitemStop [0]{}%
\providecommand \bibitemNoStop [0]{.\EOS\space}%
\providecommand \EOS [0]{\spacefactor3000\relax}%
\providecommand \BibitemShut  [1]{\csname bibitem#1\endcsname}%
\let\auto@bib@innerbib\@empty
\bibitem [{\citenamefont {Laurell}\ \emph {et~al.}(2007)\citenamefont
  {Laurell}, \citenamefont {Petersson},\ and\ \citenamefont
  {Nilsson}}]{Laurell2007}%
  \BibitemOpen
  \bibfield  {author} {\bibinfo {author} {\bibfnamefont {T.}~\bibnamefont
  {Laurell}}, \bibinfo {author} {\bibfnamefont {F.}~\bibnamefont {Petersson}},
  \ and\ \bibinfo {author} {\bibfnamefont {A.}~\bibnamefont {Nilsson}},\
  }\href@noop {} {\bibfield  {journal} {\bibinfo  {journal} {Chem Soc Rev}\
  }\textbf {\bibinfo {volume} {36}},\ \bibinfo {pages} {492} (\bibinfo {year}
  {2007})}\BibitemShut {NoStop}%
\bibitem [{\citenamefont {Bruus}\ \emph {et~al.}(2011)\citenamefont {Bruus},
  \citenamefont {Dual}, \citenamefont {Hawkes}, \citenamefont {Hill},
  \citenamefont {Laurell}, \citenamefont {Nilsson}, \citenamefont {Radel},
  \citenamefont {Sadhal},\ and\ \citenamefont {Wiklund}}]{Bruus2011c}%
  \BibitemOpen
  \bibfield  {author} {\bibinfo {author} {\bibfnamefont {H.}~\bibnamefont
  {Bruus}}, \bibinfo {author} {\bibfnamefont {J.}~\bibnamefont {Dual}},
  \bibinfo {author} {\bibfnamefont {J.}~\bibnamefont {Hawkes}}, \bibinfo
  {author} {\bibfnamefont {M.}~\bibnamefont {Hill}}, \bibinfo {author}
  {\bibfnamefont {T.}~\bibnamefont {Laurell}}, \bibinfo {author} {\bibfnamefont
  {J.}~\bibnamefont {Nilsson}}, \bibinfo {author} {\bibfnamefont
  {S.}~\bibnamefont {Radel}}, \bibinfo {author} {\bibfnamefont
  {S.}~\bibnamefont {Sadhal}}, \ and\ \bibinfo {author} {\bibfnamefont
  {M.}~\bibnamefont {Wiklund}},\ }\href {\doibase 10.1039/c1lc90058g}
  {\bibfield  {journal} {\bibinfo  {journal} {Lab Chip}\ }\textbf {\bibinfo
  {volume} {11}},\ \bibinfo {pages} {3579} (\bibinfo {year}
  {2011})}\BibitemShut {NoStop}%
\bibitem [{\citenamefont {Ding}\ \emph {et~al.}(2012)\citenamefont {Ding},
  \citenamefont {Lin}, \citenamefont {Kiraly}, \citenamefont {Yue},
  \citenamefont {Li}, \citenamefont {Chiang}, \citenamefont {Shi},
  \citenamefont {Benkovic},\ and\ \citenamefont {Huang}}]{Ding2012}%
  \BibitemOpen
  \bibfield  {author} {\bibinfo {author} {\bibfnamefont {X.}~\bibnamefont
  {Ding}}, \bibinfo {author} {\bibfnamefont {S.-C.~S.}\ \bibnamefont {Lin}},
  \bibinfo {author} {\bibfnamefont {B.}~\bibnamefont {Kiraly}}, \bibinfo
  {author} {\bibfnamefont {H.}~\bibnamefont {Yue}}, \bibinfo {author}
  {\bibfnamefont {S.}~\bibnamefont {Li}}, \bibinfo {author} {\bibfnamefont
  {I.-K.}\ \bibnamefont {Chiang}}, \bibinfo {author} {\bibfnamefont
  {J.}~\bibnamefont {Shi}}, \bibinfo {author} {\bibfnamefont {S.~J.}\
  \bibnamefont {Benkovic}}, \ and\ \bibinfo {author} {\bibfnamefont {T.~J.}\
  \bibnamefont {Huang}},\ }\href {\doibase 10.1073/pnas.1209288109} {\bibfield
  {journal} {\bibinfo  {journal} {PNAS}\ }\textbf {\bibinfo {volume} {109}},\
  \bibinfo {pages} {11105} (\bibinfo {year} {2012})}\BibitemShut {NoStop}%
\bibitem [{\citenamefont {Barnkob}\ \emph {et~al.}(2012)\citenamefont
  {Barnkob}, \citenamefont {Augustsson}, \citenamefont {Laurell},\ and\
  \citenamefont {Bruus}}]{Barnkob2012a}%
  \BibitemOpen
  \bibfield  {author} {\bibinfo {author} {\bibfnamefont {R.}~\bibnamefont
  {Barnkob}}, \bibinfo {author} {\bibfnamefont {P.}~\bibnamefont {Augustsson}},
  \bibinfo {author} {\bibfnamefont {T.}~\bibnamefont {Laurell}}, \ and\
  \bibinfo {author} {\bibfnamefont {H.}~\bibnamefont {Bruus}},\ }\href
  {\doibase 10.1103/PhysRevE.86.056307} {\bibfield  {journal} {\bibinfo
  {journal} {Phys Rev E}\ }\textbf {\bibinfo {volume} {86}},\ \bibinfo {pages}
  {056307} (\bibinfo {year} {2012})}\BibitemShut {NoStop}%
\bibitem [{\citenamefont {Doinikov}(1997)}]{Doinikov1997}%
  \BibitemOpen
  \bibfield  {author} {\bibinfo {author} {\bibfnamefont {A.~A.}\ \bibnamefont
  {Doinikov}},\ }\href {\doibase 10.1121/1.418036} {\bibfield  {journal}
  {\bibinfo  {journal} {J Acoust Soc Am}\ }\textbf {\bibinfo {volume} {101}},\
  \bibinfo {pages} {722} (\bibinfo {year} {1997})}\BibitemShut {NoStop}%
\bibitem [{\citenamefont {Danilov}\ and\ \citenamefont
  {Mironov}(2000)}]{Danilov2000}%
  \BibitemOpen
  \bibfield  {author} {\bibinfo {author} {\bibfnamefont {S.~D.}\ \bibnamefont
  {Danilov}}\ and\ \bibinfo {author} {\bibfnamefont {M.~A.}\ \bibnamefont
  {Mironov}},\ }\href {\doibase 10.1121/1.428346} {\bibfield  {journal}
  {\bibinfo  {journal} {J Acoust Soc Am}\ }\textbf {\bibinfo {volume} {107}},\
  \bibinfo {pages} {143} (\bibinfo {year} {2000})}\BibitemShut {NoStop}%
\bibitem [{\citenamefont {Settnes}\ and\ \citenamefont
  {Bruus}(2012)}]{Settnes2012}%
  \BibitemOpen
  \bibfield  {author} {\bibinfo {author} {\bibfnamefont {M.}~\bibnamefont
  {Settnes}}\ and\ \bibinfo {author} {\bibfnamefont {H.}~\bibnamefont
  {Bruus}},\ }\href {\doibase 10.1103/PhysRevE.85.016327} {\bibfield  {journal}
  {\bibinfo  {journal} {Phys Rev E}\ }\textbf {\bibinfo {volume} {85}},\
  \bibinfo {pages} {016327} (\bibinfo {year} {2012})}\BibitemShut {NoStop}%
\bibitem [{\citenamefont {King}(1934)}]{King1934}%
  \BibitemOpen
  \bibfield  {author} {\bibinfo {author} {\bibfnamefont {L.~V.}\ \bibnamefont
  {King}},\ }\href@noop {} {\bibfield  {journal} {\bibinfo  {journal} {P Roy
  Soc Lond A Mat}\ }\textbf {\bibinfo {volume} {147}},\ \bibinfo {pages} {212}
  (\bibinfo {year} {1934})}\BibitemShut {NoStop}%
\bibitem [{\citenamefont {Yosioka}\ and\ \citenamefont
  {Kawasima}(1955)}]{Yosioka1955}%
  \BibitemOpen
  \bibfield  {author} {\bibinfo {author} {\bibfnamefont {K.}~\bibnamefont
  {Yosioka}}\ and\ \bibinfo {author} {\bibfnamefont {Y.}~\bibnamefont
  {Kawasima}},\ }\href@noop {} {\bibfield  {journal} {\bibinfo  {journal}
  {Acustica}\ }\textbf {\bibinfo {volume} {5}},\ \bibinfo {pages} {167}
  (\bibinfo {year} {1955})}\BibitemShut {NoStop}%
\bibitem [{\citenamefont {Gorkov}(1962)}]{Gorkov1962}%
  \BibitemOpen
  \bibfield  {author} {\bibinfo {author} {\bibfnamefont {L.~P.}\ \bibnamefont
  {Gorkov}},\ }\href@noop {} {\bibfield  {journal} {\bibinfo  {journal} {Soviet
  Physics - Doklady}\ }\textbf {\bibinfo {volume} {6}},\ \bibinfo {pages} {773}
  (\bibinfo {year} {1962})}\BibitemShut {NoStop}%
\bibitem [{\citenamefont {Bjerknes}(1906)}]{Bjerknes1906}%
  \BibitemOpen
  \bibfield  {author} {\bibinfo {author} {\bibfnamefont {V.~F.~K.}\
  \bibnamefont {Bjerknes}},\ }\href@noop {} {\emph {\bibinfo {title} {Fields of
  Force}}}\ (\bibinfo  {publisher} {Columbia University},\ \bibinfo {year}
  {1906})\BibitemShut {NoStop}%
\bibitem [{\citenamefont {K\"onig}(1891)}]{Konig1891}%
  \BibitemOpen
  \bibfield  {author} {\bibinfo {author} {\bibfnamefont {W.}~\bibnamefont
  {K\"onig}},\ }\href@noop {} {\bibfield  {journal} {\bibinfo  {journal} {Ann.
  Phys.}\ }\textbf {\bibinfo {volume} {42}},\ \bibinfo {pages} {549} (\bibinfo
  {year} {1891})}\BibitemShut {NoStop}%
\bibitem [{\citenamefont {Emblenton}(1962)}]{Emblenton1962}%
  \BibitemOpen
  \bibfield  {author} {\bibinfo {author} {\bibfnamefont {T.~F.~W.}\
  \bibnamefont {Emblenton}},\ }\href@noop {} {\bibfield  {journal} {\bibinfo
  {journal} {J. Acoust. Soc. Am.}\ }\textbf {\bibinfo {volume} {34}},\ \bibinfo
  {pages} {1714} (\bibinfo {year} {1962})}\BibitemShut {NoStop}%
\bibitem [{\citenamefont {Nyborg}(1989)}]{Nyborg1989}%
  \BibitemOpen
  \bibfield  {author} {\bibinfo {author} {\bibfnamefont {W.~L.}\ \bibnamefont
  {Nyborg}},\ }\href@noop {} {\bibfield  {journal} {\bibinfo  {journal}
  {Ultras. Med. Biol.}\ }\textbf {\bibinfo {volume} {15}},\ \bibinfo {pages}
  {93} (\bibinfo {year} {1989})}\BibitemShut {NoStop}%
\bibitem [{\citenamefont {Doinikov}\ and\ \citenamefont
  {Zavtrak}(1995)}]{Doinikov1995}%
  \BibitemOpen
  \bibfield  {author} {\bibinfo {author} {\bibfnamefont {A.~A.}\ \bibnamefont
  {Doinikov}}\ and\ \bibinfo {author} {\bibfnamefont {S.~T.}\ \bibnamefont
  {Zavtrak}},\ }\href@noop {} {\bibfield  {journal} {\bibinfo  {journal} {Phys.
  Fluids}\ }\textbf {\bibinfo {volume} {7}},\ \bibinfo {pages} {1923} (\bibinfo
  {year} {1995})}\BibitemShut {NoStop}%
\bibitem [{\citenamefont {Crum}(1975)}]{Crum1975}%
  \BibitemOpen
  \bibfield  {author} {\bibinfo {author} {\bibfnamefont {L.}~\bibnamefont
  {Crum}},\ }\href {\doibase 10.1121/1.380614} {\bibfield  {journal} {\bibinfo
  {journal} {J Acoust Soc Am}\ }\textbf {\bibinfo {volume} {57}},\ \bibinfo
  {pages} {1363} (\bibinfo {year} {1975})}\BibitemShut {NoStop}%
\bibitem [{\citenamefont {Doinikov}\ and\ \citenamefont
  {Zavtrak}(1996)}]{Doinikov1996}%
  \BibitemOpen
  \bibfield  {author} {\bibinfo {author} {\bibfnamefont {A.~A.}\ \bibnamefont
  {Doinikov}}\ and\ \bibinfo {author} {\bibfnamefont {S.~T.}\ \bibnamefont
  {Zavtrak}},\ }\href@noop {} {\bibfield  {journal} {\bibinfo  {journal}
  {Ultrasonics}\ }\textbf {\bibinfo {volume} {34}},\ \bibinfo {pages} {807}
  (\bibinfo {year} {1996})}\BibitemShut {NoStop}%
\bibitem [{\citenamefont {Doinikov}(1996)}]{Doinikov1996a}%
  \BibitemOpen
  \bibfield  {author} {\bibinfo {author} {\bibfnamefont {A.~A.}\ \bibnamefont
  {Doinikov}},\ }\href@noop {} {\bibfield  {journal} {\bibinfo  {journal} {J.
  Acoust. Soc. Am.}\ }\textbf {\bibinfo {volume} {99}},\ \bibinfo {pages}
  {3373} (\bibinfo {year} {1996})}\BibitemShut {NoStop}%
\bibitem [{\citenamefont {Zhuk}(1985)}]{Zhuk1985}%
  \BibitemOpen
  \bibfield  {author} {\bibinfo {author} {\bibfnamefont {A.~P.}\ \bibnamefont
  {Zhuk}},\ }\href@noop {} {\bibfield  {journal} {\bibinfo  {journal} {Sov.
  Appl. Mech.}\ }\textbf {\bibinfo {volume} {21}},\ \bibinfo {pages} {110}
  (\bibinfo {year} {1985})}\BibitemShut {NoStop}%
\bibitem [{\citenamefont {Doinikov}(1999)}]{Doinikov1999}%
  \BibitemOpen
  \bibfield  {author} {\bibinfo {author} {\bibfnamefont {A.~A.}\ \bibnamefont
  {Doinikov}},\ }\href@noop {} {\bibfield  {journal} {\bibinfo  {journal} {J.
  Acoust. Soc. Am.}\ }\textbf {\bibinfo {volume} {106}},\ \bibinfo {pages}
  {3305} (\bibinfo {year} {1999})}\BibitemShut {NoStop}%
\bibitem [{\citenamefont {Doinikov}(2002)}]{Doinikov2002}%
  \BibitemOpen
  \bibfield  {author} {\bibinfo {author} {\bibfnamefont {A.~A.}\ \bibnamefont
  {Doinikov}},\ }\href@noop {} {\bibfield  {journal} {\bibinfo  {journal} {J.
  Acoust. Soc. Am.}\ }\textbf {\bibinfo {volume} {111}},\ \bibinfo {pages}
  {1602} (\bibinfo {year} {2002})}\BibitemShut {NoStop}%
\bibitem [{\citenamefont {Zheng}\ and\ \citenamefont
  {Apfel}(1995)}]{Zheng1995}%
  \BibitemOpen
  \bibfield  {author} {\bibinfo {author} {\bibfnamefont {X.}~\bibnamefont
  {Zheng}}\ and\ \bibinfo {author} {\bibfnamefont {R.~E.}\ \bibnamefont
  {Apfel}},\ }\href@noop {} {\bibfield  {journal} {\bibinfo  {journal} {J.
  Acoust. Soc. Am.}\ }\textbf {\bibinfo {volume} {97}},\ \bibinfo {pages}
  {2218} (\bibinfo {year} {1995})}\BibitemShut {NoStop}%
\bibitem [{\citenamefont {Doinikov}(2001)}]{Doinikov2001}%
  \BibitemOpen
  \bibfield  {author} {\bibinfo {author} {\bibfnamefont {A.~A.}\ \bibnamefont
  {Doinikov}},\ }\href@noop {} {\bibfield  {journal} {\bibinfo  {journal} {J.
  Fluid Mech.}\ }\textbf {\bibinfo {volume} {444}},\ \bibinfo {pages} {1}
  (\bibinfo {year} {2001})}\BibitemShut {NoStop}%
\bibitem [{\citenamefont {Morse}\ and\ \citenamefont
  {Ingard}(1986)}]{Morse1986}%
  \BibitemOpen
  \bibfield  {author} {\bibinfo {author} {\bibfnamefont {P.~M.}\ \bibnamefont
  {Morse}}\ and\ \bibinfo {author} {\bibfnamefont {K.~U.}\ \bibnamefont
  {Ingard}},\ }\href@noop {} {\emph {\bibinfo {title} {Theoretical
  Acoustics}}}\ (\bibinfo  {publisher} {Princeton University Press},\ \bibinfo
  {address} {Princeton NJ},\ \bibinfo {year} {1986})\BibitemShut {NoStop}%
\bibitem [{\citenamefont {Pierce}(1989)}]{Pierce1989}%
  \BibitemOpen
  \bibfield  {author} {\bibinfo {author} {\bibfnamefont {A.~D.}\ \bibnamefont
  {Pierce}},\ }\href@noop {} {\emph {\bibinfo {title} {Acoustics}}}\ (\bibinfo
  {publisher} {Acoustical Society of America},\ \bibinfo {address} {Melville
  NY},\ \bibinfo {year} {1989})\BibitemShut {NoStop}%
\bibitem [{\citenamefont {Blackstock}(2000)}]{Blackstock2000}%
  \BibitemOpen
  \bibfield  {author} {\bibinfo {author} {\bibfnamefont {D.~T.}\ \bibnamefont
  {Blackstock}},\ }\href@noop {} {\emph {\bibinfo {title} {Physical
  acoustics}}}\ (\bibinfo  {publisher} {John Wiley and Sons},\ \bibinfo
  {address} {Hoboken NJ},\ \bibinfo {year} {2000})\BibitemShut {NoStop}%
\bibitem [{\citenamefont {Landau}\ and\ \citenamefont
  {Lifshitz}(1993)}]{Landau1993}%
  \BibitemOpen
  \bibfield  {author} {\bibinfo {author} {\bibfnamefont {L.~D.}\ \bibnamefont
  {Landau}}\ and\ \bibinfo {author} {\bibfnamefont {E.~M.}\ \bibnamefont
  {Lifshitz}},\ }\href@noop {} {\emph {\bibinfo {title} {Fluid Mechanics}}},\
  \bibinfo {edition} {2nd}\ ed.,\ Vol.\ \bibinfo {volume} {6, Course of
  Theoretical Physics}\ (\bibinfo  {publisher} {Pergamon Press},\ \bibinfo
  {address} {Oxford},\ \bibinfo {year} {1993})\BibitemShut {NoStop}%
\bibitem [{\citenamefont {{H. \"{U}berall}}(1998)}]{Crocker1998}%
  \BibitemOpen
  \bibfield  {author} {\bibinfo {author} {\bibnamefont {{H. \"{U}berall}}},\
  }\href@noop {} {\emph {\bibinfo {title} {Handbook of Acoustics}}},\ edited
  by\ \bibinfo {editor} {\bibfnamefont {M.~J.}\ \bibnamefont {Crocker}}\
  (\bibinfo  {publisher} {John Wiley and Sons},\ \bibinfo {address} {Hoboken
  NJ},\ \bibinfo {year} {1998})\BibitemShut {NoStop}%
\bibitem [{\citenamefont {Kino}(1987)}]{kino1987}%
  \BibitemOpen
  \bibfield  {author} {\bibinfo {author} {\bibfnamefont {G.~S.}\ \bibnamefont
  {Kino}},\ }\href@noop {} {\emph {\bibinfo {title} {Acoustic Waves: Devices,
  Imaging, and Analog Signal Processing}}}\ (\bibinfo  {publisher}
  {Prentice-Hall},\ \bibinfo {address} {Upper Saddle River NJ},\ \bibinfo
  {year} {1987})\BibitemShut {NoStop}%
\bibitem [{\citenamefont {Barnkob}\ \emph {et~al.}(2010)\citenamefont
  {Barnkob}, \citenamefont {Augustsson}, \citenamefont {Laurell},\ and\
  \citenamefont {Bruus}}]{Barnkob2010}%
  \BibitemOpen
  \bibfield  {author} {\bibinfo {author} {\bibfnamefont {R.}~\bibnamefont
  {Barnkob}}, \bibinfo {author} {\bibfnamefont {P.}~\bibnamefont {Augustsson}},
  \bibinfo {author} {\bibfnamefont {T.}~\bibnamefont {Laurell}}, \ and\
  \bibinfo {author} {\bibfnamefont {H.}~\bibnamefont {Bruus}},\ }\href
  {\doibase 10.1039/b920376a} {\bibfield  {journal} {\bibinfo  {journal} {Lab
  Chip}\ }\textbf {\bibinfo {volume} {10}},\ \bibinfo {pages} {563} (\bibinfo
  {year} {2010})}\BibitemShut {NoStop}%
\bibitem [{\citenamefont {{Wolfram Research Inc.}}(2010)}]{Mathematica}%
  \BibitemOpen
  \bibfield  {author} {\bibinfo {author} {\bibnamefont {{Wolfram Research
  Inc.}}},\ }\href@noop {} {\emph {\bibinfo {title} {Mathematica 8.0}}}\
  (\bibinfo  {publisher} {Wolfram Research Inc.},\ \bibinfo {address}
  {Champaign IL},\ \bibinfo {year} {2010})\BibitemShut {NoStop}%
\bibitem [{\citenamefont {Coupland}\ and\ \citenamefont
  {McClements}(1997)}]{coupland1997}%
  \BibitemOpen
  \bibfield  {author} {\bibinfo {author} {\bibfnamefont {J.~N.}\ \bibnamefont
  {Coupland}}\ and\ \bibinfo {author} {\bibfnamefont {D.~J.}\ \bibnamefont
  {McClements}},\ }\href@noop {} {\bibfield  {journal} {\bibinfo  {journal} {J.
  Am. Oil Chem. Soc.}\ }\textbf {\bibinfo {volume} {12}},\ \bibinfo {pages}
  {1559} (\bibinfo {year} {1997})}\BibitemShut {NoStop}%
\bibitem [{\citenamefont {Nii}\ \emph {et~al.}(2009)\citenamefont {Nii},
  \citenamefont {Kikumoto},\ and\ \citenamefont {Tokuyama}}]{nii2009}%
  \BibitemOpen
  \bibfield  {author} {\bibinfo {author} {\bibfnamefont {S.}~\bibnamefont
  {Nii}}, \bibinfo {author} {\bibfnamefont {S.}~\bibnamefont {Kikumoto}}, \
  and\ \bibinfo {author} {\bibfnamefont {H.}~\bibnamefont {Tokuyama}},\
  }\href@noop {} {\bibfield  {journal} {\bibinfo  {journal} {Ultras Sonochem}\
  }\textbf {\bibinfo {volume} {16}},\ \bibinfo {pages} {145} (\bibinfo {year}
  {2009})}\BibitemShut {NoStop}%
\bibitem [{\citenamefont {Augustsson}\ \emph {et~al.}(2011)\citenamefont
  {Augustsson}, \citenamefont {Barnkob}, \citenamefont {Wereley}, \citenamefont
  {Bruus},\ and\ \citenamefont {Laurell}}]{Augustsson2011}%
  \BibitemOpen
  \bibfield  {author} {\bibinfo {author} {\bibfnamefont {P.}~\bibnamefont
  {Augustsson}}, \bibinfo {author} {\bibfnamefont {R.}~\bibnamefont {Barnkob}},
  \bibinfo {author} {\bibfnamefont {S.~T.}\ \bibnamefont {Wereley}}, \bibinfo
  {author} {\bibfnamefont {H.}~\bibnamefont {Bruus}}, \ and\ \bibinfo {author}
  {\bibfnamefont {T.}~\bibnamefont {Laurell}},\ }\href {\doibase
  10.1039/c1lc20637k} {\bibfield  {journal} {\bibinfo  {journal} {Lab Chip}\
  }\textbf {\bibinfo {volume} {11}},\ \bibinfo {pages} {4152} (\bibinfo {year}
  {2011})}\BibitemShut {NoStop}%
\bibitem [{\citenamefont {Hammarstr\"om}\ \emph {et~al.}(2014)\citenamefont
  {Hammarstr\"om}, \citenamefont {Evander}, \citenamefont {Wahlstr\"om},\ and\
  \citenamefont {Nilsson}}]{Hammarstrom2014}%
  \BibitemOpen
  \bibfield  {author} {\bibinfo {author} {\bibfnamefont {B.}~\bibnamefont
  {Hammarstr\"om}}, \bibinfo {author} {\bibfnamefont {M.}~\bibnamefont
  {Evander}}, \bibinfo {author} {\bibfnamefont {J.}~\bibnamefont
  {Wahlstr\"om}}, \ and\ \bibinfo {author} {\bibfnamefont {J.}~\bibnamefont
  {Nilsson}},\ }\href {\doibase 10.1039/C3LC51144H} {\bibfield  {journal}
  {\bibinfo  {journal} {Lab Chip}\ }\textbf {\bibinfo {volume} {14}},\ \bibinfo
  {pages} {1005} (\bibinfo {year} {2014})}\BibitemShut {NoStop}%
\end{thebibliography}

%

\end{document}